\documentclass[]{emulateapj}
\usepackage{apjfonts}

\usepackage{epsfig, natbib, graphicx, color, amsmath, amssymb}

\newcommand{\Om}{\Omega_m}

\newcommand{\be}{\begin{equation}}
\newcommand{\ee}{\end{equation}}

\newcommand{\bea}{\begin{eqnarray}}
\newcommand{\eea}{\end{eqnarray}}

\def\ba#1\ea{\begin{align}#1\end{align}}
\newcommand{\refeq}[1]{Eq.~(\ref{eq:#1})}          
\newcommand{\refeqs}[2]{Eqs.~(\ref{eq:#1})--(\ref{eq:#2})}          
\newcommand{\reffig}[1]{Fig.~\ref{fig:#1}}          
\newcommand{\vs}{\nonumber\\}       
\newcommand{\refsec}[1]{Sec.~\ref{sec:#1}}          
\newcommand{\reftab}[1]{Tab.~\ref{tab:#1}}          

\renewcommand{\S}{\Sigma}

\newcommand{\<}{\langle}
\renewcommand{\>}{\rangle}

\renewcommand{\k}{\kappa}
\newcommand{\s}{\sigma}
\renewcommand{\d}{\delta}
\newcommand{\D}{\Delta}
\newcommand{\eps}{\varepsilon}

\newcommand{\g}{\gamma}

\newcommand{\xbar}{\overline{x}}
\newcommand{\mbar}{\overline{m}}
\newcommand{\etat}{\tilde\eta}
\newcommand{\qt}{\tilde q}

\shortauthors{Schmidt et al}
\shorttitle{Magnification}

\begin{document}
\title{A Detection of Weak Lensing Magnification using Galaxy Sizes and Magnitudes}

\author{Fabian Schmidt\altaffilmark{1}}

\author{Alexie Leauthaud\altaffilmark{2}}

\author{Richard Massey\altaffilmark{3}}

\author{Jason~Rhodes\altaffilmark{1,4}}

\author{Matthew R. George\altaffilmark{5}}

\author{Anton~M.~Koekemoer\altaffilmark{6}}

\author{Alexis~Finoguenov\altaffilmark{7,8}}

\author{Masayuki~Tanaka\altaffilmark{9}}

\altaffiltext{1}{California Institute of Technology, MC 350-17, 1200
  East California Boulevard, Pasadena, CA 91125, USA}

\altaffiltext{2}{Institute for the Physics and Mathematics of the Universe, University of Tokyo, Kashiwa 2778582, Japan}

\altaffiltext{3}{Institute for Astronomy, Blackford Hill, Edinburgh
  EH9 3HJ UK}

\altaffiltext{4}{Jet Propulsion Laboratory, California Institute of Technology, Pasadena, CA 91109, USA}

\altaffiltext{5}{Department of Astronomy, University of California,
  Berkeley, CA 94720, USA}

\altaffiltext{6}{Space Telescope Science Institute, 3700 San Martin Drive, 
Baltimore, MD 21218, USA}

\altaffiltext{7}{Max-Planck-Institut f\"ur extraterrestrische Physik, 
Giessenbachstra{\ss}e, 85748 Garching, Germany}

\altaffiltext{8}{Center for Space Science Technology, University of Maryland 
Baltimore County,
1000 Hilltop Circle, Baltimore, MD 21250, USA}

\begin{abstract}
Weak lensing is commonly measured using shear through galaxy ellipticities,
or using the effect of magnification bias on galaxy number densities.  
Here, we report on the first detection of weak lensing magnification with
a new, independent technique using the distribution of galaxy sizes and 
magnitudes.  These data come for free in galaxy surveys designed for measuring
shear.  We present the magnification estimator and apply it to
an X-ray selected sample of galaxy groups in the COSMOS HST survey.  
The measurement of the projected surface density $\Sigma(r)$ is consistent
with the shear measurements within the uncertainties,
and has roughly 40\% of the signal-to-noise
of the latter.  We discuss systematic issues and challenges to realizing
the potential of this new probe of weak lensing.  
\end{abstract}

\setcounter{footnote}{0}

\section{Introduction}
\label{sec:intro}

Weak lensing has emerged as a powerful tool in cosmology and
astrophysics (e.g., \cite{Refregier,MunshiEtal,HoekstraJain,MasseyKitching,SchrabbackEtal}).  
It probes the underlying total matter distribution,
and can be used to tie the observed distribution of galaxies to
that of the dark matter, the dominant component of the cosmic
matter budget.  So far, weak lensing has primarily been measured
using the shear, through the statistical correlation of galaxy
ellipticities (see \cite{BartelmannSchneider} for a review).  
In addition, weak lensing magnification has been
detected through its effect on the number density of a flux-limited 
sample (magnification bias; 
\cite{ScrantonEtal05,HildebrandtEtal09}).   
Shear $\g$ and convergence $\k$, which are related to the
magnification $\mu$ through $\mu = [(1-\kappa^2) + |\gamma|^2]^{-1}$,
so that $\mu = 1+2\kappa$ to linear order,
measure different properties of the
density field.  For an azimuthally symmetric lens,
\ba
\g =\:& \frac{\D\Sigma}{\Sigma_{\rm cr}};\quad
\k = \frac{\Sigma}{\Sigma_{\rm cr}}\\
\Sigma_{\rm cr} =\:& \frac{c^2}{4\pi G}\frac{D_s}{D_L\,D_{Ls}} \label{eq:Scr}\\
\D\Sigma =\:& \bar\S(< r) - \S(r),
\ea
where $\Sigma(r)$ is the projected surface mass density of the lens,
$r$ is the physical transverse distance on the lens plane, and
$D_s,\:D_L,\:D_{Ls}$ denote the angular diameter distance out to the source,
lens, and between lens and source, respectively.  Note that the shear
is proportional to the differential surface mass density $\D\S$, while 
the convergence probes $\S(r)$ itself.  
The different radial dependences of $\k$ and $\g$ can
be used to break degeneracies between the lens mass and density profile shape
\citep[][]{RozoSchmidt}.  

The convergence $\k$ has previously been measured using the observed
number density of background sources.   In a
flux-limited survey, the observed number density is given by
\be
n_{\rm obs} = \bar n [ 1 + \d_g + (5s-2) \k],
\ee
where $\bar n$ is the mean number density, $\d_g$ denotes the intrinsic
fluctuation in galaxy density, and 
$s = \partial\ln\bar n/\partial m_{\rm cut}$ for a sharp magnitude 
cut $m_{\rm cut}$.  This can be generalized to surveys that are also
limited by a size cut \citep[][]{schmidtetal09}.  By cross-correlating
foreground lenses with background sources widely separated in redshift,
one can isolate this effect to measure $\k$.  Note
that the intrinsic fluctuations $\d_g$ act as noise, and any residual
physical correlation with the lenses (e.g., due to photo-z uncertainties)
will contaminate the signal.  

In this paper, we report on a detection of magnification
around X-ray selected galaxy groups in COSMOS using galaxy sizes 
and fluxes.  Related approaches have been proposed in 
\cite{jain02,BartelmannNarayan,BertinLombardi}; see also \cite{MenardEtal10} who 
used individual quasar magnitudes, and \cite{HuffGraves}.
We rely on the observed properties of individual
galaxies and do not use the number density of galaxies as done in
magnification bias measurements.  
One key advantage of this technique is that 
the contamination due to residual physical correlation with the lenses 
is likely much smaller than in magnification bias measurements,
due to the weaker correlation of galaxy sizes and luminosities 
with environment (\cite{MaltbyEtal,NairEtal,CrotonEtal}).

Throughout, we adopt a
flat $\Lambda$CDM cosmology with $h=0.72,\; \Om = 0.258$ to calculate
distances.  Halo masses are defined through a mean interior density of 
$200\rho_{\rm crit}(z=0.38)$, where $z=0.38$ is the mean redshift of the
lens sample.  

\section{Magnification Estimator}
\label{sec:est}

The data set we use consists of a set of values $\{d_i,m_i,z_i\}$
for each galaxy, denoting the size, magnitude, and photometric redshift.  
Let us consider how these observables change under lensing magnification.  
Throughout, we will work in the weak lensing regime.  We can then write
\ba
d =\:& (1 + \eta\:\k) d_0\label{eq:dlens}\\
m =\:& m_0 + q \k\label{eq:mlens}\\
z =\:& z_0.
\ea
Here, a subscript $0$ denotes unlensed quantities, and in the last
line we have assumed that photometric redshifts are not affected by lensing.  
In absence of noise and instrumental effects, the efficiency factors 
are $\eta = 1$ and
$q = -5/\ln 10 \approx -2.17$, respectively (e.g., \cite{schmidtetal09}).  
In practice, these values can be different
due to the instrument PSF, and other effects and in general depend
on the properties of a given galaxy.  

In order to estimate $\k$ using \refeqs{dlens}{mlens}, we need to know
the unlensed quantities $d_0, m_0$.  These have some intrinsic distribution, 
and given the known distribution we can infer $\k$ statistically.  Our
starting point for the estimator is thus the joint distribution
$p_0(d_0,m_0|z)$ of unlensed galaxy sizes and magnitudes at fixed redshift;
the estimator is designed to \emph{not} use the observed number density
of galaxies.  We 
approximate $p_0$ as Gaussian in $m$ and $x \equiv \ln d$.  Furthermore, 
we approximate the error distributions in $m$ and $x$ as Gaussian.  This
greatly simplifies the evaluation of the estimator, at the price
of making it sub-optimal.  
The log-likelihood is then given by
\ba
-2 \ln P\left(\S\right) =\:& \sum_i \frac{1}{1-\rho^2(z_i)} 
\label{eq:LL}\\
 \times& \Bigg\{ \frac{1}{\s_{x,i}^2} \left(x_i - \xbar(z_i) - \etat_i F_i\S\right)^2
+ \frac{1}{\s_{m,i}^2}\left(m_i - \mbar(z_i) - \tilde q_i F_i\S\right)^2\vs
- & 2 \frac{\rho(z_i)}{\s_{x,i}\s_{m,i}} \left(x_i-\xbar(z_i)-\etat_iF_i\S\right)\
\left(m_i - \mbar(z_i) - \tilde q_i F_i\S\right) \Bigg\} \vs
\s_{x,i}^2 =\:& \s_{x,{\rm meas},i}^2 + \s_{x,\rm int}^2(z_i) + \s_{xz,i}^2\\
\s_{m,i}^2 =\:& \s_{m,{\rm meas},i}^2 + \s_{m,\rm int}^2(z_i) + \s_{mz,i}^2\\
\etat_i =\:& \eta(d_i) + \eps_x(z_i)\\
\tilde q_i =\:& q(m_i) + \eps_m(z_i)\\
F_i =\:& F(z_i) = \Sigma^{-1}_{\rm cr}(z_i).
\ea
The total variance in $x$ and $m$ is given by the sum in quadrature of
the measurement uncertainty, intrinsic dispersion,
and propagated photometric redshift uncertainty.  
$\s_{x,\rm meas},\;\s_{m,\rm meas}$ are measured as a function of 
$d$ and $m$, respectively, through multiple images of the same galaxy
from overlap regions of the HST ACS data \citep[][]{cosmoswl}.  
The propagated redshift uncertainty is approximated as ($a = x,\;m$)
\be
\s_{az,i} = \frac{1}{2} |\bar a(z_{+,i})-\bar a(z_{-,i})|,
\ee
where $z_{\pm,i}$ are the upper and lower 68\% confidence level values for the redshift
of galaxy $i$.  The corrections $\eps_x,\;\eps_m$ to $\etat$ and $\qt$ take into
account the fact that a positive magnification pushes faint, small galaxies over 
the flux and size thresholds (\emph{lensing bias}).  A similar effect exists
for shear, but only at second order \citep[][]{schmidtetal09b}, while 
the magnification estimator is affected at leading order.  
The correlation between intrinsic galaxy sizes and magnitudes is
quantified by the correlation coefficient $\rho$.  

The maximum-likelihood estimate $\hat\Sigma$ and its estimated variance are then given
by
\be
\hat\Sigma = \frac{\sum_i  A_i\,F_i}{\sum_i  B_i\,F_i^2}\;;\quad
\s^2(\hat\Sigma) = \left(\sum_i  B_i\,F_i^2\right)^{-1},
\ee
where
\ba
A_i \equiv\:& \frac{1}{1-\rho^2(z_i)}\Bigg\{
\frac{\etat_i}{\s_{x,i}^2} (x_i - \xbar(z_i))
+ \frac{\tilde q_i}{\s_{m,i}^2} (m_i - \mbar(z_i))\vs
&\hspace*{1.6cm}
- \frac{\rho(z_i)}{\s_{x,i}\s_{m,i}} \left[\etat_i (m_i - \mbar(z_i))
+ \tilde q_i (x_i - \xbar(z_i)) \right] \Bigg \}\\
B_i \equiv\:& \frac{1}{1-\rho^2(z_i)}\left\{
\frac{\etat_i^2}{\s_{x,i}^2} + \frac{\tilde q_i^2}{\s_{m,i}^2}
- 2 \frac{\rho(z_i)}{\s_{x,i} \s_{m,i}}\etat_i \tilde q_i \right\}.
\ea
The intrinsic dispersions $\s_{x,\rm int},\, \s_{m,\rm int}$ as well as 
$\rho$ are determined by fitting a bivariate Gaussian
distribution to all galaxies before cuts on $x,\,m$ in a
redshift slice.  We subtract the mean measurement error in 
$x,\,m$ in quadrature from the total
measured dispersion to obtain $\s_{x,\rm int},\,\s_{m,\rm int}$.  

The quantities $\xbar,\,\mbar$ are determined by solving the equation
$\hat\Sigma=0$ applied to all galaxies in a given redshift slice.  
Since $F\approx\rm const$ within a narrow redshift slice, this yields
\ba
&\sum_i \frac{1}{1-\rho^2(z_i)}\left\{
\frac{\etat_i}{\s_{x,i}^2} (x_i - \xbar(z_i))
- \frac{\rho(z_i)}{\s_{x,i}\s_{m,i}} \tilde q_i (x_i - \xbar(z_i)) \right\}
= 0\\
& \sum_i \frac{1}{1-\rho^2(z_i)}\left\{
 \frac{\tilde q_i}{\s_{m,i}^2} (m_i - \mbar(z_i))
- \frac{\rho(z_i)}{\s_{x,i}\s_{m,i}} \etat_i (m_i - \mbar(z_i)) \right\} = 0.
\ea
$\xbar,\,\mbar$ do not correspond to the true mean intrinsic quantities, 
since only galaxies passing cuts are used to measure them.  We can then write
\ba
\bar a(z) =\:& \left(\sum_i w_{a,i}\right)^{-1} 
\sum_i w_{a,i}\: a_i\label{eq:xbar},\\
 w_{x,i} =\:& \frac{1}{1-\rho^2(z_i)}\left\{
\frac{\etat_i}{\s_{x,i}^2} - \frac{\rho(z_i)}{\s_{x,i}\s_{m,i}}\tilde q_i \right\}\\
 w_{m,i} =\:& \frac{1}{1-\rho^2(z_i)}\left\{
\frac{\tilde q_i}{\s_{m,i}^2} - \frac{\rho(z_i)}{\s_{x,i}\s_{m,i}}\etat_i \right\},
\ea
where the sum runs over all galaxies passing cuts in the redshift slice.  

In order to determine $\eps_x,\,\eps_m$, we repeat the measurement of $\xbar,\,\mbar$
with magnitude and size cuts varied by $\pm\Delta m,\;\pm\D x$, respectively,
where $\D m = q(m_{\rm cut}) \k_0$ and $\D x = \eta(d_{\rm cut}) \k_0$,
and we take $\k_0=0.02$ (the value of $\eps_i$ does not depend
significantly on $\k_0$).  Then, 
\ba
\eps_x(z) =\:& \frac{\xbar_+(z) - \xbar_-(z)}{2\k_0}\\
\eps_m(z) =\:& \frac{\mbar_+(z) - \mbar_-(z)}{2\k_0},
\ea
where $\bar a_{\pm}$ indicate quantities measured when using
cuts $m_{\rm cut}\pm \D m$ and $x_{\rm cut}\pm\D x$.  In order to take into
account interdependencies, we iterate the measurement
of $\xbar(z),\,\mbar(z)$ until convergence.   

\begin{figure}[t]
\centering
\includegraphics[width=0.49\textwidth,clip=true,trim=0cm 0.3cm 0cm 0cm]
{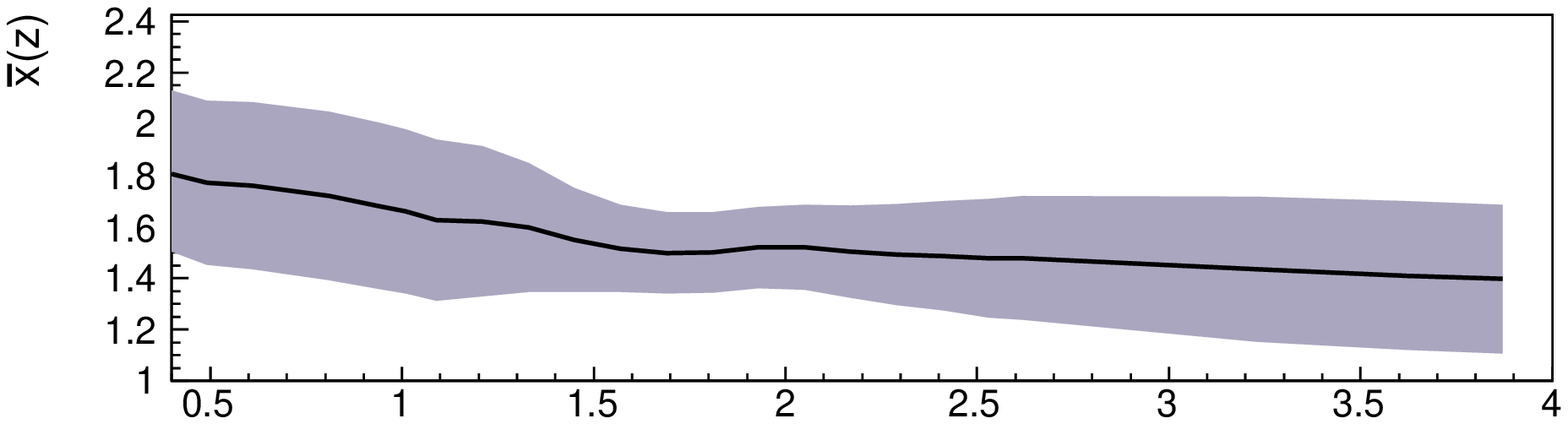}
\includegraphics[width=0.49\textwidth,clip=true,trim=0cm 0cm 0cm 0cm]
{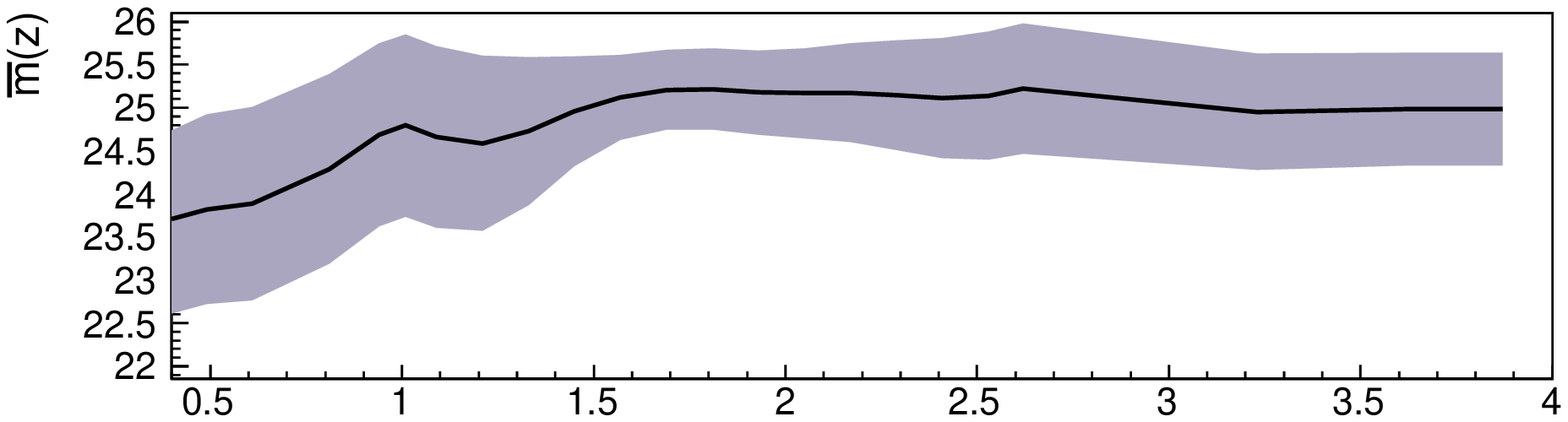}

\caption{\textit{Upper panel:}  Average source galaxy log size $\xbar(z)$ as 
a function of redshift, measured using \refeq{xbar}.  The shaded band
shows $\s_{x,\rm int}$.   
\textit{Lower panel:}  Same, for source galaxy magnitudes ($\mbar(z)$, 
$\sigma_{m,\rm int}$).
\label{fig:xbar}}
\end{figure}

\section{Data}
\label{sec:data}

Our data set consists of the HST ACS imaging of the COSMOS field
\citep[][]{Koekemoeretal},
from which we take the galaxy sizes $d$ (in pixels of 0.03'') and 
magnitudes $m$ 
in the F814W band.  For the size $d$ we use the variable-width Gaussian-filtered
second moment of \cite{RRG}, while magnitudes $m$ are \texttt{MAG\_AUTO} 
from SExtractor \citep[][]{Sextractor}.   We leave more sophisticated, optimized 
size and flux estimators for future work.  
The details of the reduction pipeline are described in \cite{cosmoswl}.  The
size estimator is PSF-deconvolved, and we thus expect systematic effects
of PSF variations to be negligible.   In order to avoid any contamination
by a background-dependence of the estimated sizes, we only
use sky regions within the lower peak (mean pixel background $< 0.0062$)
of the essentially bimodal background 
distribution in the COSMOS ACS data, comprising $\sim 75$\% of the field
(see Fig.~8 in \cite{cosmoswl}).  

For our source galaxy sample, we exclude galaxies in masked regions
\citep[][]{IlbertEtal} and apply the cuts on 
size, magnitude and redshift summarized in \reftab{cuts}.  The size cut 
approximately
corresponds to a measured size (before PSF deconvolution) of 1.4 times
the PSF size.  Note that
in the shear analysis, a significantly stricter cut ($d>3.6$) is employed.  
The magnitude cut is mostly determined by the reliability of the photometric 
redshifts.  Finally, the redshift cut excludes a small portion of the sample 
which does not contribute significantly to the lensing signal-to-noise.  
We obtain a sample of $N_{\rm src}=250,500$ galaxies with a median 
redshift of 1.22.  In the lensing analysis, we also impose the conditions 
$z_{i,-} > z_L$, $z_i > z_L+0.2$ for each source-lens pair.  

\begin{figure}[t]
\centering
\includegraphics[width=0.49\textwidth,clip=true,trim=0cm 0.15cm 0cm 0cm]{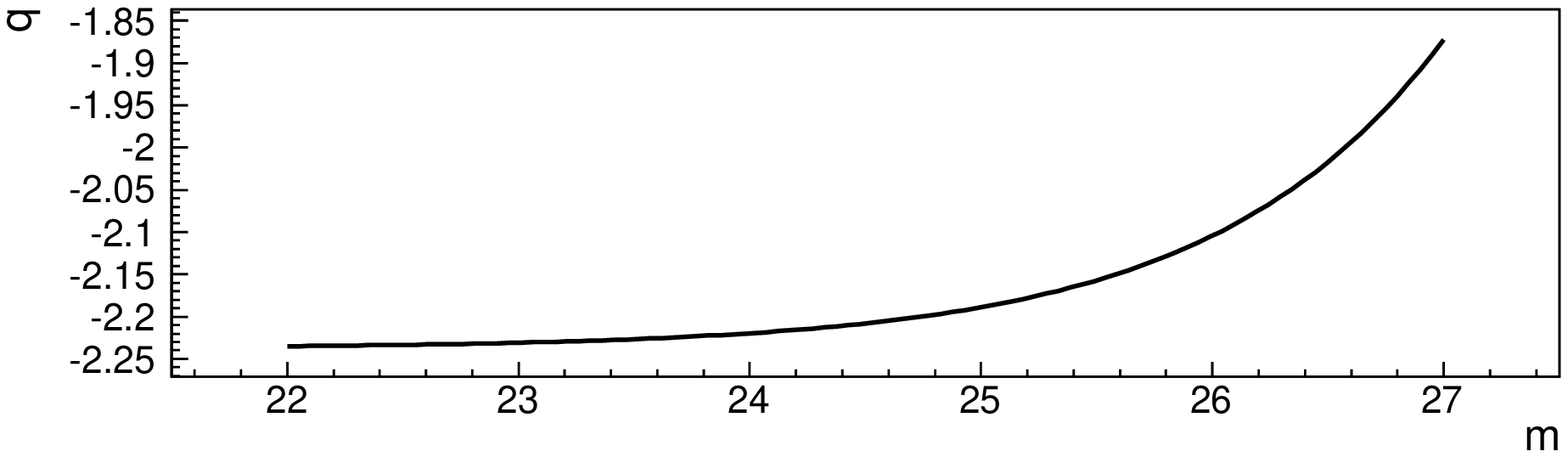}
\includegraphics[width=0.49\textwidth,clip=true,trim=0cm 0.3cm 0cm 0.1cm]
{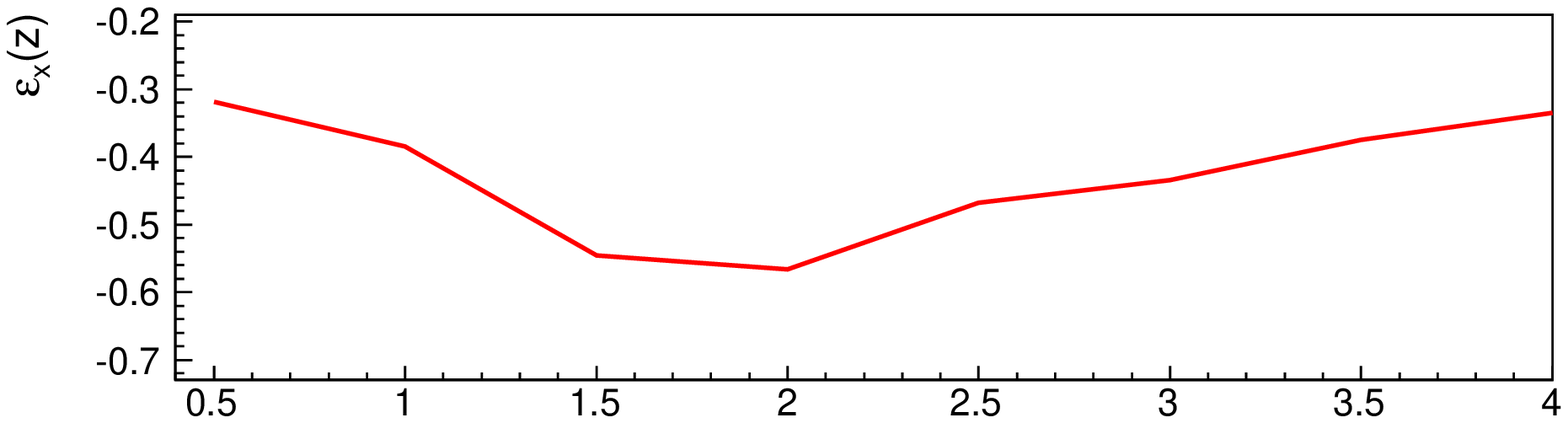}
\includegraphics[width=0.49\textwidth,clip=true,trim=0cm 0cm 0cm 0cm]
{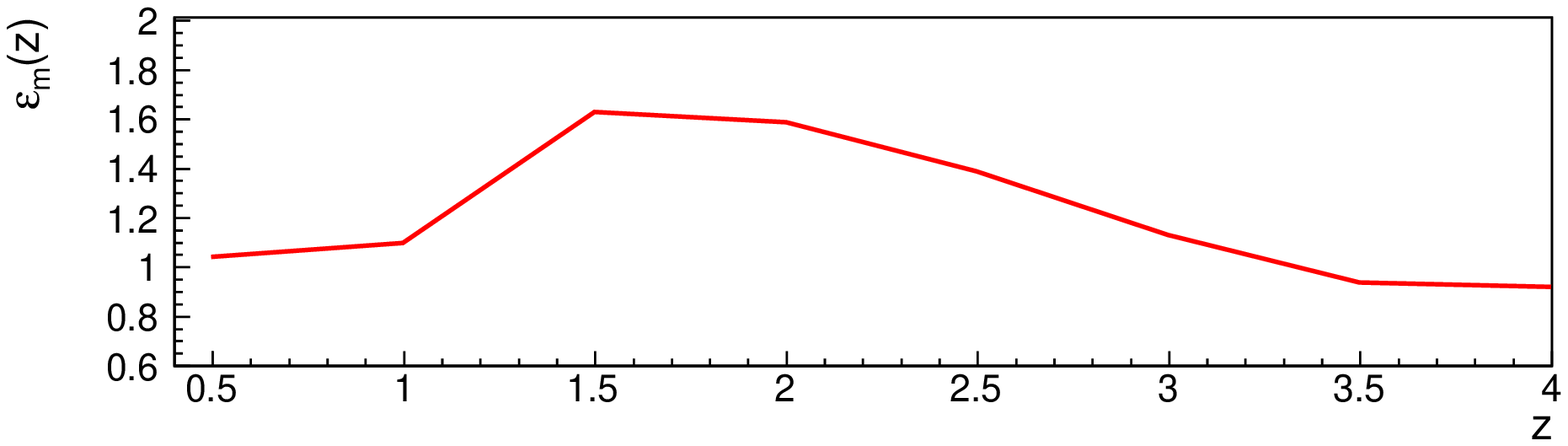}
\caption{\textit{From top to bottom:~a)} Flux lensing efficiency $q$
as a function of magnitude $m$.  \textit{b)}  Lensing bias correction $\eps_x$ 
for galaxy (log) sizes.  The effective size lensing
efficiency is given by $\etat = \eta + \eps_x$.  \textit{c)} 
Lensing bias effect $\eps_m$ for  magnitudes.  Correspondingly,
$\qt = q + \eps_m$.
\label{fig:eff}}
\end{figure}

We apply the fitting procedure described in the last section,
using overlapping redshift slices of half-width $\D z = 0.25$
separated by $\D z$.   \reffig{xbar} illustrates the 
measured $\xbar(z),\; \mbar(z)$ as well as $\s_{x,\rm int},\; \s_{m,\rm int}$.  
We fix $\rho=-0.6$ to improve the stability of the fit.  Allowing
$\rho$ to vary does not significantly improve the fit.
\reffig{eff} shows $\eps_x$ and $\eps_m$.  
Note that $\eps_x < 0$: the number of faint, small galaxies
passing the cuts for $\k > 0$ reduces the average size, counteracting
the lensing effect on the sizes themselves.   The analogous holds
for magnitudes, where $\eps_m > 0$.  This effect can in principle
be avoided by cutting on a quantity not affected by lensing (e.g., surface 
brightness) instead of magnitude and size.  We leave the exploration
of this for future work.  

\begin{table}[b]
\centering
\caption{Summary of cuts applied to the source galaxy sample.}
\label{tab:cuts}
\begin{tabular}{l|c}
\hline
$m$ & $< 25.8$~mag \\
$d$ & $ > 2$~pixels \\
$z$ & $\in [0.5,\;4]$ \\
\hline
\end{tabular}
\end{table}

While the estimators in \refsec{est} were designed to rely as much
as possible on data alone, we do need simulations to determine
the lensing efficiencies $\eta,\;q$.  
To simulate ACS data, we use the {\tt simage} software package 
\citep[][]{simage1,simage2,simage3}, 
as  previously adopted for the Shear TEsting Programme \citep[STEP;][]{step2} 
and to calibrate the COSMOS shear analysis \citep[][]{cosmoswl}.  
The parameters were tuned to mimic the galaxy source counts  and 
correlated background noise properties of the reduced and stacked 
COSMOS images.  We model the intrinsic morphologies of source galaxies via 
shapelets \citep[][]{shapelets1,shapelets2}.  
We have adapted {\tt simage} to enable the superposition 
of an input convergence $\k_{\mathrm{sim}}$ and produced images at 15 values 
of $\k_{\mathrm{sim}}\in[-0.2,0.3]$.  

To determine how each simulated galaxy responds to magnification,
we match galaxies in the lensed and 
unlensed ($\k_{\rm sim}=0$) simulations.  
We measure the average $x$ and $m$ as a function of the
input $\k_{\rm sim} = -0.1\dots 0.1$.  
Then, we estimate $\eta$ through
\be
\eta = \frac{\<x\>_{\k_{\rm sim}=\k} - \<x\>_{\k_{\rm sim}=0}}{\k},
\ee
and correspondingly for $q$.  We have found that for galaxies passing cuts,
$\eta$ is within $[0.9,1.1]$ and consistent with 1
for the whole sample (recall that we use a PSF-deconvolved size estimator).  

For the lens sample, we use X-ray selected groups since these have been well 
studied using shear \citep[][]{groups}.  
The entire COSMOS region has been mapped by {\sl XMM-Newton}, while the central 
region is covered by {\sl Chandra} observations \citep[][]{HasingerEtal07,CappellutiEtal09,ElvisEtal09}.  These
X-ray data have been used to construct a group catalog
containing 211 extended X-ray sources, 
spanning the range $0<z_{\rm gr}<1$ \citep[][]{Finoguenov:2007,groups}.
Group membership is assigned to galaxies in the COSMOS catalog
based on the photometric redshifts and the red sequence.  
We include only groups with reliable optical associations which have
more than 3 members, and we exclude close neighboring systems and
groups near the edges of the field or masked regions 
(\textsc{flag\_include}=1 in \cite{GeorgeEtal11a}).   We use the location
of the most massive group galaxy (MMGG) located within the NFW scale radius
of the X-ray centroid as the group center
(MMGG$_{\rm scale}$ as defined in \cite{GeorgeEtal11b}).  
\cite{GeorgeEtal11b} show that this
is likely the most reliable estimate for the group centers.   
We restrict the redshift range of groups to 0.2-0.6, close to the peak of 
the lensing sensitivity of the source sample.  This selection also reduces the 
potential impact of catastrophic redshift errors which become more prevalent at 
$z_s > 1.5$, and yields $N_{\rm gr} = 61$ groups.

\begin{figure}[t]
\centering
\includegraphics[width=0.49\textwidth]{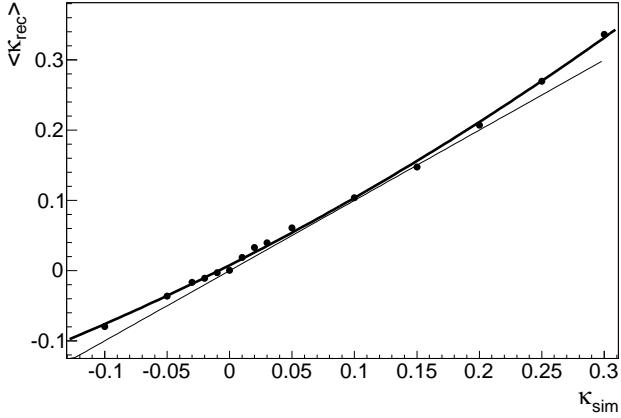}
\caption{Mean reconstructed convergence $\k$ in simulations with magnification,
as a function of the input $\k_{\rm sim}$.  The thin solid line shows the identity while the thick
solid line shows a quadratic fit.
\label{fig:MC}}
\end{figure}

\section{Results}
\label{sec:results}

Before applying our estimators to data, we performed the calibration
described in \refsec{est} on the unlensed simulations, and applied
the estimators to the lensed simulations as a consistency check.  
Since the simulations do not contain redshift information, we
set $F(z_i) = 1$.  
\reffig{MC} shows the mean reconstructed $\k$ as a function of the
true $\k_{\rm sim}$, along with a quadratic fit,
\be
\<\k_{\rm rec}\> = p_0 + p_1 \k_{\rm sim} + p_2 \k_{\rm sim}^2.
\ee
Note that we expect some quadratic correction in this relation, since 
our estimators neglect all second- and
higher-order terms.  We found $p_1=0.95\pm0.05$, indicating that
our estimator is close to unbiased, and $p_2\approx 0.6$.  

Turning to data, 
we collect source galaxies in bins of physical transverse radius $r$
around the $N_{\rm gr}$ X-ray groups, and apply the estimator for
$\Sigma$.  The second order corrections amount to at most $\sim 10$\% at the 
smallest radii, as shown in \cite{groups}, and can be neglected 
given the errors of this measurement.  
Since $\xbar(z),\:\mbar(z)$ are measured in finite redshift bins without any
redshift weighting while we apply redshift weighting in the $\Sigma$
measurement, we need to subtract a constant offset $\S_{\rm off}$ from the 
measurement.  $\Sigma_{\rm off}$ is measured by pairing all galaxies 
passing cuts in the COSMOS field with each group, yielding $\Sigma_{\rm off} = 151.9\:M_\odot/\rm pc^2$.  
This is equivalent to estimating $\<\S(r)\>_{\rm groups} - \<\S\>_{\rm COSMOS}$, 
where both averages employ the same redshift weighting.  
Note that a circle of $r\sim 1$~Mpc drawn around each group 
covers a significant fraction of the COSMOS field; thus, we necessarily
subtract some of the signal in this way.  
If we instead restrict the source sample used for the $\S_{\rm off}$ estimate to
galaxies separated by at least 1~Mpc in physical transverse distance from
each group in the sample, $\S_{\rm off}$ decreases by $4.4\:M_\odot/\rm pc^2$,
thus increasing $\S(r)$ uniformly by this amount.  

\begin{figure}[t]
\centering
\includegraphics[width=0.49\textwidth]{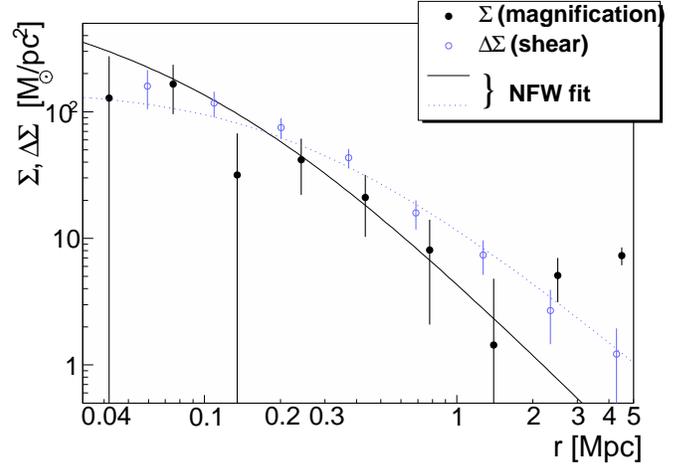}
\caption{Projected surface mass density $\Sigma$ (filled circles) measured 
around groups using the estimator described in \refsec{est}.  
The open circles show $\D\S$ measured using shear for the same group sample.  
The solid and dotted lines show $\Sigma$ and $\D\S$, respectively, for the 
joint best-fitting NFW model ($\lg (M/M_{\odot}) = 13.43$; $r < 1$~Mpc).  
Note that associated
large-scale structure contributes significantly to $\S(r)$ for
$r \gtrsim 1$~Mpc.  
\label{fig:Sigma}}
\end{figure}

The measured $\S(r)$ is shown in \reffig{Sigma}.  
The error bars are those returned by the Gaussian estimators.  
Measurements of the variance of $\hat\Sigma$
applied to random sets of source galaxies indicate that these errors
are correct to within $\sim$15\%.  
For $r \lesssim 1$~Mpc, the errors for different radial bins are independent.  

We fit an NFW profile \citep[][]{NFW} with fixed concentration $c=4$ to the measurement, 
calculating $\S(r)$ as described e.g. in \cite{SchmidtRozo}.  
Furthermore, we restrict the fitting to $r < 1$~Mpc, 
beyond which the contribution from associated large-scale structure
becomes important (see below).  
This yields a best-fit mass of $\lg (M/M_\odot) = 13.25\pm 0.16$
(statistical error), corresponding to a detection significance of $\sim 3.8\s$.  
This corresponds to $\sim 38$\% of the signal-to-noise of shear.  
Note that the precise value also depends on the definition
of the offset $\S_{\rm off}$.  
The best-fit mass is consistent with the estimate from shear, 
$\lg (M/M_\odot) = 13.49\pm 0.07$ (\cite{groups}; 
again for $r < 1$~Mpc).   

In principle, a measurement of $\Sigma$ in addition to $\D\S$ can break
the mass-concentration degeneracy, since the two quantities 
have different radial dependencies \citep[][]{RozoSchmidt}.  Due to the 
limited signal-to-noise of this first measurement, a joint fit to
the $\Sigma$ measurement together with $\D\S$ from shear only yields an improvement
of  $\sim 10$\% in the area of the error ellipse in the mass-concentration plane.  
The $\Sigma$ measurements at $r > 1$~Mpc are consistent with an estimate of 
the associated large-scale structure contribution (two-halo term) using the 
halo model, which yields $\S_{2h} \sim 10\:M_\odot/\rm pc^2$ with
weak $r$-dependence for $r < 5$~Mpc.  The corresponding contribution to 
the shear is much smaller, $\D\S_{2h} \lesssim 1\:M_\odot/\rm pc^2$, 
further exemplifying the 
complementarity of the two measurements.  A detailed modeling of these 
data will be the subject of forthcoming work.

The main systematic uncertainty of the $\S$ measurement is 
due to the lensing bias effect.  
We have quantified the impact of the measurement errors 
on $\eps_x,\:\eps_m$ by varying $m_{\rm cut},\:x_{\rm cut}$, resulting
in an overall multiplicative systematic error on $\Sigma$ of 20\%.  
Residual effects from PSF and background variations have been found 
to be negligible.  Increasing the minimum size cut from 2 to 3 pixels
yields 
compatible results, as do estimators using sizes and magnitudes separately.

The systematic uncertainty due to photometric
redshift errors is similar to that of the shear (\cite{groups};
note that we have employed the same source-lens
separation criteria as in that paper).  Given the lower signal-to-noise 
of the magnification measurement, this systematic is negligible.

\section{Summary and Discussion}
\label{sec:concl}

We have presented the first measurement of weak lensing magnification
using galaxy sizes and fluxes.  Measurements of galaxy sizes and
magnitudes come for free with weak lensing
surveys.  We find a signal-to-noise only a factor of $\sim 2-3$ less than
shear;  thus it is worth pursuing this kind of independent weak
lensing measurement.  Magnification constitutes a different observable 
than shear, and the non-local relationship between the two
can be used to break degeneracies e.g. when measuring halo profiles.  

Furthermore, the systematics affecting this measurement are
largely independent from those of the shear:  to first order, sizes are
only sensitive to the size of the PSF, while magnitudes are PSF-independent.  
On the other hand, a knowledge of the shape of the PSF is necessary
for shear measurements.  This allows us to use less restrictive cuts.  
Conversely, a careful calibration of the lensing bias correction
is necessary for the magnification measurement, while this effect is
not present in the shear at linear order \citep[][]{schmidtetal09b}.
Nevertheless, given sufficiently 
detailed simulations, this is a solvable problem.

While the measurement presented here is limited by signal-to-noise
and statistics, it is clear that for future, much larger surveys,
robust size and flux estimators are essential to a successful application
of this method.  Given the already promising results, we believe
a dedicated effort to develop such estimators for weak lensing 
will be worthwhile.  Finally, our results show that there is significant
lensing information in photometric galaxy samples in addition to the
shear alone;  this should be taken into consideration
in the design of future weak lensing experiments.

\acknowledgements
We thank Jessica~Ford, Hendrik~Hildebrandt, Eric~Huff,
Bhuvnesh~Jain, Donghui~Jeong, 
Eric~Jullo, James~Taylor, and Ludovic~van~Waerbeke for helpful discussions.  
FS is supported by the Gordon and Betty Moore Foundation.  
RM is supported by an STFC Advanced Fellowship.  JR was supported by JPL, 
run by Caltech for NASA.



\end{document}